\documentclass[12pt]{article}
\usepackage{amsfonts}
\usepackage{amsmath}
\usepackage{mathrsfs}
\usepackage{color}
\usepackage{amssymb}
\newcommand{\bea}{\begin{eqnarray}}
\newcommand{\eea}{\end{eqnarray}}
\newcommand{\be}{\begin{equation}}
\newcommand{\ee}{\end{equation}}
\newcommand{\vs}[1]{\vspace{#1 mm}}

\renewcommand{\a}{\alpha}
\renewcommand{\b}{\beta}
\renewcommand{\c}{\gamma}
\renewcommand{\d}{\delta}
\newcommand{\e}{\epsilon}
\newcommand{\dsl}{\pa \kern-0.5em /}

\newcommand{\pa}{\partial}

\newcommand{\nn}{\nonumber\\}

\def\ii{{\rm i}}

\begin{document}

\topmargin 0pt \oddsidemargin 0mm

\begin{flushright}

USTC-ICTS-09-15\\

\end{flushright}

\vspace{2mm}

\begin{center}

{\Large \bf The long-range interactions between branes \\in diverse
dimensions}

\vs{6}

{\large Rong-Jun Wu \footnote{E-mail: rjwu@mail.ustc.edu.cn} and
Zhao-Long Wang \footnote{E-mail: zlwang4@mail.ustc.edu.cn}}

\vspace{6mm}

{\em

Interdisciplinary Center for Theoretical Study\\

University of Science and Technology of China, Hefei, Anhui 230026, China\\

}

\end{center}

\begin{abstract}

We calculate the long-range interactions between two simple branes
placed parallel at a separation in diverse dimensions via an
effective field theory approach. We also compute for the first time
the explicit long-range interaction between two D-branes with each
carrying a world-volume non-abelian magnetic flux in three special
cases, respectively. In particular, we demonstrate that the
half-string creation between a D$_0$-brane and a D$_8$-brane continues to hold even in
the present context, therefore lending further support to the
previous assertion of this. Our computations re-raise also the issue
in one case on whether so constructed (D$_0$, D$_8$) bound state is
actually a marginal one.

\end{abstract}
\newpage

\section{Introduction}

Usually, we have three methods to calculate the static interaction
between two branes placed parallel at a separation: the stringy
computations \cite{Polchinski:1995aa, Polchinski:1996aa} if a string
description is applicable, the brane-probe computations
\cite{Dabholkar:1990yf, Lu:1994dk, Tseytlin:1996aa, Tseytlin:1997c}
and the effective field theory computations \cite{Polchinski:1996aa,
DiVecchia:1999uf, Lu:2009nwx}. Each method has its own
applicability, advantage and disadvantage. For the string-level
computations and for D-branes, the lowest-order stringy interaction
can be computed either as an open string one-loop annulus diagram
with one end of the open string located at one D-brane and the other
end at the other D-brane or as a closed string tree-level cylinder
diagram with one D-brane emitting a closed string, propagating for a
certain amount of time and finally absorbed by the other D-brane.
The validity of this computation requires a small string coupling.
For the brane-probe computations, we consider a brane probe moving
in a background produced by a brane source. Obviously, the probe
must not change the background, i.e., the number of the probe branes
should be much smaller than that of the source branes. The
interaction can be obtained by finding the potential of the probe in
the fields of source brane. While in the effective field theory
computations, once the effective field theories both in the bulk and
on the world-volume are given, we can find the propagator for each
bulk (massless) mode and the corresponding coupling with the brane,
and the interaction between branes can be calculated subsequently,
for example, following \cite{Lu:2009nwx}. Here, we need neither
brane to be heavier than the other nor the explicit configuration of
the source brane. The first method has its advantage if the branes
such as D-branes have a stringy description and the string coupling
is small. The second and third methods may also be good if only the
low-energy effective descriptions both for the bulk and for the
branes are available such as the case for the M2-brane and the M5-brane in
M-theory. In particular, for the (transverse) M5-brane, the explicit
interaction between two such branes can be calculated only this way
at present. For the second approach, one needs in addition the
explicit configuration of the source brane which may not be always
available. So the least requirement is for the third approach and
the interaction computed should be good for large brane separation
in general, sometime even at stringy level such as the one between a
D$_0$-brane and a D$_8$-brane. This is the focus of the present paper for brane
interactions in diverse dimensions.

In this paper, we calculate the long-range Coulomb-type interaction
between two $p$-branes placed parallel at a separation via the
effective field theory computations mentioned above. The $p$-branes
in diverse dimensions follow the brane-scan given in
\cite{Duff:1992hu, Lu:1994dk}. For finding the interaction, only the
bosonic part of the bulk or the brane world-volume effective action
is needed. For cases where the world-volume modes involve a vector
such as for a D-brane or a tensor such as for a M5-brane, we will
set them vanish since these modes will in general not be excited for
calculating the long-range interaction. Then the bosonic part of the
effective world-volume action coupled with the bulk modes can be
simply described by the usual Nambu-Goto type action plus the
corresponding Wess-Zumino term, involved only the scalars in the
respective multiplet which are described by the spacetime embedding
coordinates. This kind of computations serves to check the
``no-force" condition beyond the probe approach at large brane
separation for BPS $p$-branes, sometime even valid at stringy level
such as the case for the D$_0$-D$_8$ system. For certain cases such as for
M2-brane, M5-brane and NS5-brane in type IIA where a better description such as a
stringy one is not available in general, this computation is
particularly useful. The long-range interaction between a $p$-brane
and an anti $p$-brane can be obtained from the above by switching
the sign of the contribution due to the $(p + 1)$-form potential.

When specified to D-branes, we can consider cases with various
constant world-volume fluxes. The effective action for the branes is
now the usual DBI action plus the corresponding Wess-Zumino term or
the non-abelian extension of this. Given the bulk fields as
described by the low energy type II supergravities, the relevant
couplings are all gauge singlets even when the effective brane
action is non-abelian. When there is one abelian flux present, this
has been considered in \cite{Lu:2009nwx}.  In this paper, we will
consider three different cases with each involving a special
world-volume non-abelian magnetic flux in the spirit of
\cite{Taylor:1997aa}, which corresponds to either (D$_{p - 4}$,
D$_p$) or (D$_{p - 6}$, D$_p$) or (D$_0$, D$_8$) configuration, and
calculate the corresponding long-range interaction. In particular,
for the last system, this computation can be even valid at the
stringy level for the contribution due to the D$_0$-brane and D$_8$-brane in the
interaction between two such bound states since it is well-known
that only the massless modes contribute to this part of interaction.
Further, for the same system, the previously discovered half-string
creation \cite{Danielsson:1997wq, Bergman:1997gf, Kitao:1998vn,
Billo:1998np} and the associated divergent zero-point energy
\cite{Lu:2009w} between a D$_0$-brane and a D$_8$-brane are found to continue to hold
even in the present case, therefore lending further support to both.
Moreover, when the half-string creation is considered, the net
interaction between two such bound states is beautifully canceled,
just like the case of (D$_{p - 4}$, D$_p$), showing the
long-believed marginal bound-state nature of (D$_0$, D$_8$). This
re-raises the issue on whether so constructed (D$_0$, D$_8$) bound
state given in \cite{Taylor:1997aa} and used in the present
calculations is actually a marginal one.

This paper is organized as follows. In Section 2, we calculate the
static long-range interaction between two  simple $p$-branes placed
parallel at a separation in diverse dimensions. We start with the
relevant bosonic part of supergravity action which is the bulk low
energy effective action and from this we can read the propagator for
each relevant bulk field when expressed in the canonical form. From
the relevant brane world-volume action coupled with bulk fields, we
can read the respective couplings. From these, the respective
long-range interaction between two simple branes can be simply
calculated in diverse dimensions. For BPS branes, we can check the
``no-force" condition beyond the probe approach. Also, the
interaction between a $p$-brane and an anti $p$-brane can also be
similarly calculated. In Section 3, we specify to D-branes with
special world-volume non-abelian  magnetic fluxes, which corresponds
to either (D$_{p - 4}$, D$_p$) or (D$_{p - 6}$, D$_p$) or (D$_0$,
D$_8$) configuration, and calculate the corresponding long-range
interaction between two such branes in type II string theories in a
similar spirit. We discuss various issues regarding the (D$_0$, D$_8$)
bound state such as the half-string creation and its nature as a
marginal bound state.  We discuss the results and conclude this
paper in Section 4.

\section{The interaction between two $p$-branes in diverse dimensions}

In this section, we will calculate the static long-range interaction
between two simple $p$-branes placed parallel at a separation in
diverse dimensions. The two $p$-branes  can be both (anti-) BPS ones
or one is BPS and the other is anti-BPS. In the former case, a
net-zero interaction is expected, while in the latter a
non-vanishing result is expected. For this, we first express the
relevant bosonic part of the bulk effective action, i.e., the bulk
supergravity action, in the canonical form in spacetime dimension $D$.
With this, we can find the couplings of the $p$-brane with the
relevant bulk (massless) fields through the corresponding
world-volume effective action which is taken as the Nambu-Goto one
plus the Wess-Zumino term. We then calculate the long-range
interaction between two such $p$-branes as described in the
Introduction in diverse dimensions.

The relevant bosonic part of supergravity in spacetime dimension $D$
with $p$-brane $\sigma$-model metric $G_{\mu\nu}$ \footnote{The
Greek indices $\mu$, $\nu$, \ldots label the spacetime directions 0,
1, \ldots, $D$.} is \cite{Duff:1993ye}
 \bea \label{pbulkb}
S_D &=& \frac{1}{2 \kappa_D ^2}\int d^D x \sqrt
{-G}e^{-\frac{(D-2)\a(p)}{2(p+1)}\Phi}\left[ R\right.\nn& &
~~~~\left.-\frac{1}{2}\left(1 -
\frac{\a^2(p)(D-1)(D-2)}{2(p+1)^2}\right)\left(\nabla
\Phi\right)^2-\frac{1}{2}|F_{p+2}|^2\right],
 \eea
where the $(p+2)$-form field strength $F_{p+2}$ is given by $F_{p+2}
= d C_{p+1}$ with $C_{p + 1}$ the $(p + 1)$-form potential, and
$\a(p)$ satisfies
 \bea \label{ap}
\a^2(p)=4-\frac{2(p+1)(D-p-3)}{D-2}.
 \eea
To consider the field theory limit, it is proper to express the
above action in the Einstein or canonical frame. This can be
achieved through the so-called Einstein metric $g_{\mu\nu}$ which is
related to the $p$-brane $\sigma$-model metric $G_{\mu\nu}$ as
 \bea\label{WelyTransfb}
g_{\mu\nu} = e^{-\frac{\a(p)}{p+1}\phi}G_{\mu\nu},
 \eea
where
 \bea\label{dilaton}
\phi \equiv  \Phi - \Phi_0
 \eea
with $\Phi_0$ the asymptotic value (or VEV ) of the
dilaton \footnote{We choose the $p$-brane $\sigma$-model metric
$G_{\mu\nu}$ to be asymptotically flat.}. In this frame,  we have
 \bea \label{pbulke}
S_D = \frac{1}{2 \kappa ^2}\int d^D x \sqrt
{-g}\left[R-\frac{1}{2}\left(\nabla
\phi\right)^2-\frac{1}{2}e^{-a(p)\phi}|F_{p+2}|^2\right].
 \eea
In the above, we introduce the physical gravitational coupling
$2\kappa^2 = 2g_b^2\kappa_D^2$ with the dimensionless parameter $g_b
= e^{\frac{(D-2)\a(p)}{4 (p+1)}\Phi_0}$, which becomes the string
coupling when the 10 $D$ fundamental string is considered.

Considering small fluctuations of fields with respect to the flat
Minkowski background $g_{\mu\nu}=\eta_{\mu\nu}+h_{\mu\nu}$ and
choosing the usual harmonic gauge for $h_{\mu\nu}$, we have the
action
 \bea \label{pbulks}
S_D = \frac{1}{2 \kappa ^2}\int d^D x \left[-\frac{1}{4}\nabla
h^{\mu\nu}\nabla h_{\mu\nu} + \frac{1}{8}(\nabla
h)^2-\frac{1}{2}\left(\nabla
\phi\right)^2-\frac{1}{2}|F_{p+2}|^2\right],
 \eea
where we keep only the lowest order terms. The above action
obviously becomes canonical with the following scalings:
 \bea \label{pcanonfield}
h_{\mu\nu} \to 2\kappa h_{\mu\nu},\,\,\,\phi \to \sqrt 2 \kappa
\phi,\,\,\,C_{p+1}\to \sqrt 2 \kappa C_{p+1}.
 \eea
These will help us to determine the corresponding couplings of bulk
fields with the $p$-brane in the canonical form which we will turn
next.

Let us consider the bosonic world-volume action of a $p$-brane
 \bea\label{pwvN}
S = -T_p \int d^{p+1}\sigma \sqrt {-G} + T_p\int C_{p+1},
 \eea
where the metric $G$ and the $(p+1)$-form potential $C_{p+1}$ are
the pullbacks of the corresponding bulk fields to the world-volume,
and $T_p$ is the $p$-brane tension. In the above, the first term is
the Nambu-Goto action and the second one is the Wess-Zumino term.
Using Eq. (\ref{WelyTransfb}), we can express the above action in
Einstein frame as
 \bea\label{pwv}
S = -T_p \int d^{p+1}\sigma e^{\a(p)\phi /2}\sqrt {-g} + T_p\int
C_{p+1},
 \eea
We expand the above action to the leading order for the same
background fluctuations  and end up with \footnote{The Greek indices
$\a$, $\b$, \ldots label the world-volume directions 0, 1, \ldots,
$p$ along which the $p$-brane extends.}
 \bea \label{pwvs}
S = -T_p \int d^{p+1}\sigma \left(1 + \frac{1}{2}\eta^{\a\b}h_{\a\b}
+ \frac{\a(p)}{2}\phi\right)+T_p\int C_{p+1}.
 \eea
Using the scalings in Eq. (\ref{pcanonfield}) to replace the
background fluctuations in the above action, we can obtain the
respective coupling in the canonical form
 \bea\label{pgc}
J_h^{(i)} = -n_ic_p V_{p+1}\eta^{\a\b}h_{\a\b}
 \eea
for the graviton,
 \bea\label{pdc}
J_\phi^{(i)} = -\frac{\a(p)}{\sqrt 2}n_ic_p V_{p+1}\phi
 \eea
for the dilaton, and
 \bea\label{pCpp1c}
 J_{C_{p+1}}^{(i)} =  \frac{\sqrt 2 n_i c_p}{(p+1)!} V_{p+1}
C_{\a_0\a_1\cdots\a_p}\e^{\a_0\a_1\cdots\a_p}
 \eea
for the $(p+1)$-form potential $C_{p+1}$. In the above, $c_p \equiv
T_p\kappa$, $V_{p+1}$ is the world-volume of the $p$-brane,
$\e^{\a_0\a_1\cdots\a_p}$ is the totally antisymmetric tensor on the
$p$-brane world-volume \footnote{By conventions, $\e^{01\cdots
p}=-\e_{01\cdots p}=1.$}, and the index $i$ denotes the respective
stack of $p$-branes with $i=1,2$. Note that we have introduced an
extra overall integral factor $n_i$ in each couplings to count the
multiplicity of $n_i$ coincident $p$-branes in each stack.

Now we calculate the lowest-order contribution in momentum
space \footnote{The corresponding potential in coordinate space can
be obtained simply by Fourier transformation following, for example,
\cite{Lu:2009nwx}.} to the interaction between two $p$-branes placed
parallel to each other at a given separation due to the exchanges of
massless modes, therefore representing the interaction at large
separation.

The gravitational potential energy density due to the exchange of
graviton is
 \bea\label{pge}
U_h = \frac{1}{V_{p+1}}\underbrace{J_h^{(1)}J_h^{(2)}} = n_1n_2
c_p^2 V_{p+1}\eta^{\a\b}\eta^{\c\d}\underbrace {h_{\a\b}h_{\c\d}},
 \eea
where the propagator is
 \bea \label{gp}
\underbrace {h_{\a\b}h_{\c\d}} =
\left[\frac{1}{2}(\eta_{\a\c}\eta_{\b\d} + \eta_{\a\d}\eta_{\b\c}) -
\frac{1}{D-2}\eta_{\a\b}\eta_{\c\d}\right]\frac{1}{k_\bot^2}
 \eea
from Eq. (\ref{pbulks}) for the canonically normalized graviton
propagating in the transverse directions, so we have
 \bea \label{pge1}
U_h = n_1n_2c_p^2 \frac{V_{p+1}}{k_\bot^2}\left[\left(p+1\right)-
\frac{\left(p+1\right)^2}{D-2}\right].
 \eea
Similarly,  the potential energy density due to the exchange of
dilaton $\phi$ and the one due to the $(p+1)$-form potential
$C_{01\cdots p}$ can be calculated, respectively, as
 \bea\label{pde}
U_\phi = \frac{1}{V_{p+1}}\underbrace{J_\phi^{(1)}J_\phi^{(2)}} =
\frac{\a^2(p)}{2}n_1n_2c_p^2 V_{p+1} \underbrace{\phi\phi}=
\frac{\a^2(p)}{2}n_1n_2c_p^2 \frac{V_{p+1}}{k_\bot^2}
 \eea
and
 \bea\label{pcpp1e}
U_{C_{p+1}} =
\frac{1}{V_{p+1}}\underbrace{J_{C_{p+1}}^{(1)}J_{C_{p+1}}^{(2)}} =
2n_1n_2 c_p^2 V_{p+1} \underbrace{C_{01\cdots p}C_{01\cdots p}}= - 2
n_1n_2 c_p^2 \frac{V_{p+1}}{k_\bot^2}.
 \eea
 In the above, we have used the respective propagator for dilaton and
for the $(p+1)$-form potential as
 \bea \label{dp}
\underbrace {\phi\phi} = \frac{1}{k_\bot^2}
 \eea
and
 \bea\label{cpp1p}
\underbrace {C_{01\cdots p}C_{01\cdots p}} = - \frac{1}{k_\bot^2}.
 \eea
So the total contribution to the energy density is
 \bea\label{pe}
U = U_h + U_\phi + U_{C_{p+1}} = n_1n_2 c_p^2
\frac{V_{p+1}}{k_\bot^2}\left[\left(p+1\right)-
\frac{\left(p+1\right)^2}{D-2} + \frac{\a^2(p)}{2} - 2\right] = 0,
 \eea
where we have used Eq. (\ref{ap}) in the last step.

From above, we know that the contributions from graviton, dilaton
and $(p+1)$-form potential cancel among themselves  exactly, so the
net interaction between $p$-branes vanishes. This is expected. It is
well-known that two parallel static BPS branes separated by a
distance feel no force between them, i.e., satisfying the
``no-force'' condition, and this configuration preserves 1/2 of
spacetime supersymmetries.

If one stack of branes in the above is replaced by the corresponding
anti-branes, the contribution from the $(p + 1)$-form potential will
switch sign and the resulting net interaction is no longer
vanishing. It is now
 \bea\label{peanti}
U =U_h + U_\phi - U_{C_{p+1}}= 4n_1n_2 c_p^2
\frac{V_{p+1}}{k_\bot^2},
 \eea
which shows that the interaction between $p$-branes and anti
$p$-branes is attractive \footnote{We choose conventions here that
$U> 0$ means attractive and $U < 0$ means repulsive which differ
from standard ones by a sign.}. This is due to that all the
components are attractive and the underlying system breaks all the
supersymmetries. Note that our computations go beyond the probe
approach for which we don't need one set of branes to be much
lighter than the other set. In addition, not every case considered
has a stringy description, for examples, the IIA NS5-brane case and
M-brane case to which we turn next.

For M-brane, i.e., M2-brane or M5-brane, the relevant bulk action is the bosonic
part of $D=11$ supergravity which has no dilaton. It is
 \bea\label{supergravity11}
S_{11}=\frac{1}{2\kappa^2}\int d^{11} x
\sqrt{-g}\left(R-\frac{1}{2}|F_4|^2\right)-\frac{1}{12\kappa^2}\int
F_4\wedge F_4 \wedge C_3,
 \eea
where $F_4$ is the 4-form field strength of the 3-form potential
$C_3$. The second term, the Chern-Simons-like term, has no play to
leading order in the small fluctuations since it is a higher order
term. In essence, the long-range force calculated using the
propagators for the relevant bulk fields (the graviton and the
3-form potential) read from the above bulk action and the
corresponding couplings read from the effective world-volume action
of M-brane coupled with these bulk fluctuations is just a special
case of the above general calculations from Eqs. (\ref{pbulkb}) -
(\ref{peanti}) when $D = 11$ and $p = 2$ or $5$ are taken. This is
due to that  $\a(p)=0$ in $D=11$ for $p = 2$ or $5$ and from Eq.
(\ref{pdc}) we have $J^{(i)}_\phi = 0$. This implies that the
dilaton decouples, therefore giving no contribution to the
interaction. In other words, the above general calculations apply
also to M2-brane or M5-brane. The above results for the case of the IIA NS5-brane
or the (transverse) M5-brane are the only known ones beyond the
probe approach.

\section{The interactions between two D-branes with non-abelian fluxes}

We now specify our discussion to D-branes. The long-range
interaction between two parallel D-branes with each carrying a
single abelian world-volume flux, which describes the non-threshold
BPS (F, D$_p$) bound state \cite{Witten:1996np, Arfaei:1998np,
Lu:1999jhep, Lu:1999np, Hashimoto:1997pr, DiVecchia:2000np} or
non-threshold BPS (D$_{p-2}$, D$_p$) bound state
\cite{Breckenridge:1996pr, Costa:1996zd, DiVecchia:1997np}, has been
discussed in {\cite{Lu:2009nwx}}. Some discussions regarding
multiple abelian fluxes have been given in \cite{Kitao:1998vn}. In
this section, we will consider the cases when the D-branes carry
special world-volume non-abelian magnetic fluxes. We will calculate
the couplings of the D-branes with the bulk massless modes of the
underlying type II theories through the corresponding world-volume
effective action and bulk effective action of a given string theory
(IIA or IIB), and use these couplings to find the long-range
interaction between two such D-brane configurations. In particular,
for the system of (D$_0$, D$_8$), we will address issues such as the
half-string creation between a D$_0$-brane and a D$_8$-brane and the associated
divergent zero-point energy in the present context.

Let us first express the bulk fields in the effective action of a
given string theory in canonical forms \footnote{This part follows
the e-print edition on arXiv of \cite{Lu:2009nwx}.} and we only
need to consider the corresponding bosonic action too. Since this
works the same way in either IIA or IIB theory, we take IIA for
illustration. The bosonic part of the IIA low-energy effective
action in string frame is
 \bea\label{bulk2a}
S_{{\rm IIA}} &=& S_{{\rm NS}} + S_{{\rm R}} + S_{{\rm CS}}, \nn
S_{{\rm NS}} &=& \frac{1}{2\kappa_{10}^2}\int d^{10} x \sqrt {-G}
e^{-2\Phi}\left[R+4(\nabla \Phi)^2-\frac{1}{2}|H_3|^2\right], \nn
S_{{\rm R}} &=& - \frac{1}{4\kappa_{10}^2}\int d^{10} x \sqrt {-G}
\left[|F_2|^2+|\tilde F_4|^2\right], \nn S_{{\rm CS}} &=& -
\frac{1}{4\kappa_{10}^2}\int B_2\wedge F_4\wedge F_4,
 \eea
where NS-NS field $H_3 = dB_2$ while the R-R fields $F_2 = dC_1$,
$\tilde F_4 = dC_3 - C_1\wedge H_3$. The constant $2\kappa_{10}^2$
appearing in the action is $ 2\kappa_{10}^2 = (2\pi)^7\a^{\prime
4}$.

To express the above action in the Einstein or canonical frame, we
introduce the Einstein metric $g_{\mu\nu}$ as
 \bea\label{WelyTransf}
g_{\mu\nu} = e^{-\phi/2}G_{\mu\nu},
 \eea
where $\phi$ is defined as in Eq. (\ref{dilaton}). In this frame, we
have
 \bea\label{bulk2aNS}
S_{{\rm NS}} &=& \frac{1}{2g_s^2\kappa_{10}^2}\int
d^{10}x\sqrt{-g}\left [R - \frac{1}{2}{(\nabla \phi)}^2 -
\frac{1}{2}e^{-\phi}|H_3|^2 \right ], \nn S_{{\rm R}} &=& -
\frac{1}{4\kappa_{10}^2}\int d^{10}x \sqrt {-g}\left
[e^{3\phi/2}|F_2|^2 + e^{\phi/2}|\tilde F_4|^2\right ],
 \eea
while the $S_{{\rm CS}}$ remains the same. In the above, we have
introduced the string coupling $g_s = e^{\Phi_0}$ and with this the
physical gravitational coupling is $2\kappa^2 =
2g_s^2\kappa_{10}^2$.

Similarly, considering small fluctuations of fields with respect to
the flat Minkowski background, we have the action
 \bea\label{bulk2as}
S_{{\rm IIA}} &=& \frac{1}{2\kappa^2}\int d^{10}x
\left[-\frac{1}{4}\nabla h^{\mu\nu}\nabla h_{\mu\nu} +
\frac{1}{8}(\nabla h)^2 - \frac{1}{2}{(\nabla \phi)}^2 -
\frac{1}{2}|H_3|^2\right] \nn & & ~~~~ -
\frac{1}{4\kappa_{10}^2}\int d^{10}x \left[|F_2|^2 + |F_4|^2\right],
 \eea
where $F_4 =dC_3$. The above action obviously becomes canonical with
the following scalings:
 \bea\label{canonfieldNS}
h_{\mu\nu} \to 2\kappa h_{\mu\nu},\,\,\, \phi \to \sqrt 2
\kappa\phi,\,\,\,B_{\mu\nu} \to \sqrt 2 \kappa B_{\mu\nu}
 \eea
for NS-NS fields and
 \bea\label{canonformR}
C_n \to \sqrt 2\kappa_{10}C_n
 \eea
for rank-$n$ R-R potential. Note that the scaling of a NS-NS field
differs from that of a R-R potential by a string coupling $g_s$
except for the graviton which has an additional factor of $\sqrt 2$.

To obtain the couplings of bulk fields with the D-branes which carry
world-volume non-abelian magnetic fluxes, we turn to consider the
bosonic world-volume action of these D-branes. Before proceeding,
let us mention a few simple facts.  It is well-known that the system
of $N$ coincident D$_p$-branes is described in the low-energy regime
by a $U(N)$ super Yang-Mills theory in $(p+1)$ dimensions, which can
be obtained via the dimensional reduction of the $D=10$ super
Yang-Mills \cite{Witten:1995aa}. In the Yang-Mills theory,
D$_{p-2k}$-branes ($p \ge 2 k$) within the D$_p$-branes can be
described by a configuration of gauge field $ F$ on the world-volume
of D$_p$-branes with their charge related to the topological charge
proportional to the integral of $ F^{\wedge k} = F\wedge \cdots
\wedge  F$ where the number of wedge products is $k - 1$ with $k$ an
integer \cite{Witten:1995ab, Douglas:1995aa}. In what follows, we
will consider three special world-volume non-abelian  magnetic
fluxes following \cite{Taylor:1997aa} with $k = 2$, $k = 3$ and $k =
4$. They correspond to D$_{p-4}$-branes within D$_p$-branes,
D$_{p-6}$-branes within D$_p$-branes and D$_0$-branes within
D$_8$-branes, respectively.

The bosonic world-volume action of  D$_p$-branes with a constant
non-abelian world-volume flux $ F$ in string frame is
\cite{Myers:1999jhep}
 \bea\label{fwv} S &=& -T_p \int d^{p+1}\sigma ~{\rm
Tr}\left\{e^{-\Phi}\sqrt {-{\rm det}(G+B+\hat F)}\right\} \nn & &
~~~~~~~~~~~~+ T_p \int{\rm Tr}\left\{ \left [e^{B+\hat F}\wedge
\sum_k C_{p+1-2k}\right ]_{p+1}\right\},
 \eea
where the metric $G$, the NS-NS rank-2 potential $B$,  and the R-R
potential $C_{p+1-2k}$ are the pullbacks of the corresponding bulk
fields to the world-volume. Each of these fields is a singlet under
the $U(N)$ gauge group, therefore they each can be represented by
their bulk field multiplying an $N \times N$ unit matrix $I_N$ in
the present context. In Eq. (\ref{fwv}), we define $\hat F = 2\pi
\a' F$ with $F$ an $N \times N$ matrix under the gauge group $U
(N)$, and denote `Tr' the trace in this $N \times N$ space. Note
that `$\det$' denotes the determinant with respect to the
world-volume indices only.  The subscript in the square bracket in
the above Wess-Zumino term means that in expanding the exponential
form one picks up only terms of total degree of $(p+1)$. We now
express the above action in Einstein frame using Eqs.
(\ref{WelyTransf}) and (\ref{dilaton}) as
 \bea\label{fwv1}
S &=&  -\frac{T_p}{g_s} \int d^{p+1}\sigma~{\rm Tr}\left \{
e^{\frac{(p-3)\phi}{4}}\sqrt {-{\rm det}\left[g+\left(B+\hat
F\right)e^{-\phi/2}\right]}\right \} \nn & &~~~~~~~~~~~+ T_p \int
{\rm Tr}\left \{\left [e^{B+\hat F}\wedge \sum_k C_{p+1-2k}\right
]_{p+1}\right \}.
 \eea
By the same token, we now expand the above action  to the leading
order with fixed $\hat F$ for small background fluctuations  and
have
 \bea\label{fwvs}
S &=&  - \frac{T_p}{g_s}\int d^{p+1}\sigma ~{\rm Tr} \left\{\sqrt
{-{\rm det}(\eta + \hat F)} \left [
I_N+\frac{1}{2}\left(\left(\eta+\hat
F\right)^{-1}\right)^{\a\b}\left(h_{\b\a}+B_{\b\a}\right)
\right.\right.\nn & & \,\,\,\,\,\,\,\,\,\,\,\,\,\,\,\, \left.\left.+
\frac{1}{4}\left(\left(p-3\right)I_N-{\rm tr}\left(\hat F\left(\eta
+ \hat F\right)^{-1}\right)\right)\phi\right]\right\} \nn & &
\,\,\,\,\,\,\,\,\,\,\,\,\,\,\,\,+T_p\int {\rm Tr} \left\{
C_{p+1}+\hat F\wedge C_{p-1} + \frac{1}{2!}\hat F\wedge\hat F\wedge
C_{p-3}+\cdots \right\},
 \eea
where $\cdots$ means terms with the lower rank of R-R potentials
wedged with more $\hat F$'s and the trace `tr' is  with respect to
the world-volume coordinate indices. From the above action we can
read the respective couplings in the canonical form \footnote{We
would like to point out that from now on the bulk fluctuations such
as $h_{\a\b}$, $\phi$, $B_{\b\a}$, $C_{p+1}$ and $C_{p+1-2k}$ are
just the usual ones without multiplying each with the unit matrix
$I_N$.}
 \bea\label{fgc}
J_h = -c_p V_{p+1}{\rm Tr} \left\{ \sqrt {-{\rm det}(\eta +\hat
F)}\left[\left(\eta+\hat F\right)^{-1}\right]\right\}^{\a\b}h_{\a\b}
 \eea
for the graviton,
 \bea\label{fdc}
J_\phi = \frac{c_p}{2\sqrt 2}V_{p+1}{\rm Tr} \left\{\sqrt {-{\rm
det}(\eta +\hat F)} \left[\left(3-p\right)I_N+{\rm tr}\left(\hat
F\left(\eta + \hat F\right)^{-1}\right)\right]\right\}\phi
 \eea
for the dilaton,
 \bea\label{fBc}
J_B =  -\frac{c_p}{\sqrt 2}V_{p+1}{\rm Tr}\left\{\sqrt {-{\rm
det}(\eta + \hat F)}\left[\left(\eta + \hat
F\right)^{-1}\right]\right\}^{\a\b}B_{\b\a}
 \eea
for the Kalb-Ramond field,
 \bea\label{fcpp1c}
J_{C_{p+1}} =\frac{\sqrt 2N c_p}{(p+1)!}V_{p+1}
C_{\a_0\a_1\cdots\a_p}\e^{\a_0\a_1\cdots\a_p}
 \eea
for the R-R potential $C_{p+1}$, and
 \bea\label{fcpmc}
J_{C_{p+1-2k}} = \frac{\sqrt 2 c_p}{2^k\,k!\,(p+1-2k)!}V_{p+1}{\rm
Tr} \left\{\hat F_{\a_0\a_1}\cdots \hat
F_{\a_{2k-2}\a_{2k-1}}\right\}C_{\a_{2k}\a_{2k+1}\cdots\a_p}\e^{\a_0\a_1\cdots\a_p}
 \eea
for the R-R potential $C_{p+1-2k}$. In the above, we have used $c_p
= T_p\kappa /g_s = T_p\kappa_{10}$. We will use these couplings to
calculate the long-range interactions between two D-branes with the
non-vanishing integral of $\hat F\wedge \hat F$ or $\hat F\wedge \hat F \wedge \hat F$ or
$\hat F\wedge \hat F\wedge \hat F \wedge \hat F$, i.e., the $k = 2$ or $k = 3$ or $k =
4$ case mentioned above.

\subsection{The $k = 2$ case}

When the integral of $\hat F \wedge \hat F$ is the only non-vanishing one, we
have D$_{p-4}$ (or $\bar {\rm D}_{p - 4}$)-branes within
D$_p$-branes uniformly delocalized along the flux directions.  We
can choose the  constant non-abelian magnetic flux $\hat F$ on the
world-volume of D$_p$-branes the following way
 \bea\label{flux4}
\hat F =
\begin{pmatrix} 0_{2n} & \, & \, &\, &\, &\, &\,&\, \\
\, & \cdot & \, &\, &\, &\, &\,&\,\\
\, & \, & \cdot &\, &\, &\, &\,&\,\\
\, & \, & \, &\cdot &\, &\, &\,&\,\\
\, & \, & \, &\, &0_{2n} &-f\cdot u &\,&\,\\
\, & \, & \, &\, &f\cdot u &0_{2n}  &\,&\,\\
\, & \, & \, &\, &\, &\, &0_{2n} &-f\cdot u\\
\, & \, & \, &\, &\, &\, &f\cdot u &0_{2n} \end{pmatrix}
_{(p+1)\times(p+1)},
 \eea
where $0_{2n}$ in the above matrix stands for a $2n\times 2n$ zero
matrix, and
 \bea\label{u4}
u ={\rm Diag}\{I_n,-I_n\}
 \eea
with $I_n$ the $n\times n$ unit matrix. Note that $u$ is one of the
Cartan  subalgebra generators of the $U(N)$ algebra and we consider
here $N=2n$ ($n$ is a positive integer). With this flux, it is
obvious that ${\rm Tr} (\hat F\wedge \cdots \wedge \hat F) \neq 0$
only when the number of $\hat F$ is 2 in the wedge product.
Therefore the only non-vanishing coupling associated with the lower
rank R-R potential, according to Eq. (\ref{fcpmc}), is for $k = 2$
and the corresponding R-R potential is $C_{p - 3}$. This further
implies the presence of D$_{p - 4}$-branes within D$_p$-branes whose
charge is determined by the integral of ${\rm Tr} (\hat F \wedge
\hat F)$. This brane configuration whose energy is the sum of the
energy of D$_{p-4}$-branes and D$_p$-branes is a marginally bound
state and preserves 1/4 of spacetime supersymmetries
\cite{Taylor:1997aa}. We will denote this configuration as
(D$_{p-4}$, D$_p$) in the following.

With this special flux, we have
 \bea
-{\rm det}\left(\eta + \hat F\right) = \left(1+f^2\right)^2 I_{2n},
 \eea
and
 \bea\label{tV4}
V = \left(\eta + \hat F\right)^{-1} =
\begin{pmatrix} -I_{2n} &\, &\, &\, &\, &\, &\, &\, &\,  \\
\, & I_{2n} & \, & \, &\, &\, &\, &\,&\, \\
\, & \, & \cdot & \, &\, &\, &\, &\,&\,\\
\, & \, & \, & \cdot &\, &\, &\, &\,&\,\\
\, & \, & \, & \,&\cdot &\, &\, &\,&\,\\
\, & \, & \, & \, & \,  &\frac{I_{2n}}{1+f^2} &\frac{f\cdot u}{1+f^2} &\,&\,\\
\, & \, & \, & \, & \,  &-\frac{f\cdot u}{1+f^2} &\frac{I_{2n}}{1+f^2}  &\,&\,\\
\, & \, & \, & \, & \,  &\, &\, &\frac{I_{2n}}{1+f^2} &\frac{f\cdot u}{1+f^2}\\
\, & \, & \, & \, & \,  &\, &\, &-\frac{f\cdot u}{1+f^2} &
\frac{I_{2n}}{1+f^2}
\end{pmatrix}_{(p+1)\times(p+1)}.
 \eea
So we have the couplings $J_h$, $J_\phi$, $J_B$, $J_{C_{p+1}}$ and
$J_{C_{p-3}}$ for the corresponding fields as follows,
 \bea\label{fNSc4}
J_h^{(i)} &=& -2n_i c_p V_{p+1}(1+f_i^2)\tilde V_i^{\a\b}h_{\a\b},
\nn J_\phi^{(i)} &=& -2n_i \frac{c_p}{2\sqrt 2} V_{p+1}
(1+f_i^2)\left[\left(p-3\right)-4\frac{f_i^2}{1+f_i^2}\right]\phi,
\nn J_B^{(i)} &=& -2n_i \frac{c_p}{\sqrt 2} V_{p+1}(1+f_i^2)\tilde
V_i^{\a\b}B_{\b\a}
 \eea
for the NS-NS fields and
 \bea\label{fRc4}
J_{C_{p+1}}^{(i)} &=& 2n_i \sqrt 2 c_p V_{p+1}C_{01\cdots p}, \nn
J_{C_{p-3}}^{(i)} &=& 2n_i \sqrt 2 c_p V_{p+1} f_i^2 C_{01\cdots
p-4}
 \eea
for the R-R fields. In the above,  $\tilde V_i$ is a
${(p+1)\times(p+1)}$ diagonal matrix on the world-volume of
D$_p$-branes as
 \bea\label{V4}
\tilde V_i = {\rm Diag}\{-1, \, 1,\, \cdots,\, 1,\,
\frac{1}{1+f_i^2},\, \frac{1}{1+f_i^2},\, \frac{1}{1+f_i^2},\,
\frac{1}{1+f_i^2}\},
 \eea
and this immediately implies that $J^{(i)}_B = 0$, therefore giving
zero contribution to the interaction from this coupling. In the
above, the index $i = 1, 2$, denoting the respective (D$_{p - 4}$,
D$_p$) in the interacting system.

We now use the above couplings to calculate the long-range
interaction energy density in momentum space between two parallel
(D$_{p-4}$, D$_p$) separated by a transverse distance. The
gravitational potential energy density due to the exchange of
graviton is
 \bea\label{fge4}
U_h &=& \frac{1}{V_{p+1}}\underbrace{J_h^{(1)}J_h^{(2)}} \nn &=&
c_p^2 V_{p+1} 4n_1n_2 \left(1+f_1^2\right)\left(1+f_2^2\right)\tilde
V_1^{\a\b}\tilde V_2^{\c\d}\underbrace {h_{\a\b}h_{\c\d}} \nn &=&
c_p^2
\frac{V_{p+1}}{k_\bot^2}4n_1n_2\left(\frac{-p^2+6p+7}{8}+\frac{-p^2+10p-21}{8}f_1^2\right.
\nn & & \left.\,\,\,\,\,\,\,\,\,\,\,\,\,\,\,\,+
\frac{-p^2+10p-21}{8}f_2^2+ \frac{-p^2+14p-33}{8}f_1^2 f_2^2\right),
 \eea
where in the last equality we have used the graviton propagator Eq.
(\ref{gp}). With the dilaton propagator Eq. (\ref{dp}), the
contribution to the interaction due to the exchange of dilaton can
be calculated as
 \bea\label{fde4}
U_\phi &=& \frac{1}{V_{p+1}}\underbrace{J_\phi^{(1)}J_\phi^{(2)}}
\nn &=& \frac{1}{8}c_p^2 V_{p+1}
4n_1n_2\left(1+f_1^2\right)\left(1+f_2^2\right) \nn & & ~~~~~~~~
\left[\left(p-3\right)-4\frac{f_1^2}{1+f_1^2}\right]
\left[\left(p-3\right)-4\frac{f_2^2}{1+f_2^2}\right]
\underbrace{\phi\phi} \nn &=& c_p^2 \frac{V_{p+1}}{k_\bot^2} 4n_1n_2
\left(\frac{p^2-6p+9}{8}+\frac{p^2-10p+21}{8}f_1^2 \right.\nn &
&\left.~~~~~~~~+ \frac{p^2-10p+21}{8}f_2^2+
\frac{p^2-14p+49}{8}f_1^2 f_2^2\right).
 \eea
 The total energy density from the NS-NS sector
is
 \bea\label{fNSe4}
U_{{\rm NS-NS}} = U_h + U_\phi = 8n_1n_2\left(1+f_1^2
f_2^2\right)c_p^2 \frac{V_{p+1}}{k_\bot^2},
 \eea

We now turn to the calculations of the contributions from R-R
fields. The contribution from the exchange of R-R potential
$C_{01\cdots p}$ is
 \bea\label{fcpp1e4}
U_{C_{p+1}} &=&
\frac{1}{V_{p+1}}\underbrace{J_{C_{p+1}}^{(1)}J_{C_{p+1}}^{(2)} }\nn
&=& 2 c_p^2 V_{p+1} 4n_1n_2\underbrace{C_{01\cdots p}C_{01\cdots
p}}\nn &=& -8n_1n_2 c_p^2 \frac{V_{p+1}}{k_\bot^2} ,
 \eea
where the rank-$(p+1)$ R-R potential propagator is Eq.
(\ref{cpp1p}). Similarly we have
 \bea\label{fcpm3e4}
U_{C_{p-3}} &=&
\frac{1}{V_{p+1}}\underbrace{J_{C_{p-3}}^{(1)}J_{C_{p-3}}^{(2)}} \nn
&=& 2 c_p^2 V_{p+1} 4n_1n_2 f_1^2f_2^2 \underbrace{C_{01\cdots
p-4}C_{01\cdots p-4}}\nn &=& -8n_1n_2f_1^2f_2^2 c_p^2
\frac{V_{p+1}}{k_\bot^2},
 \eea
where the propagator for the rank-$(p-3)$ R-R potential has the same
form as Eq. (\ref{cpp1p}), i.e.,
 \bea\label{cpm3p}
\underbrace {C_{01\cdots p-4}C_{01\cdots p-4}} = -
\frac{1}{k_\bot^2}.
 \eea
So the total energy density from the R-R sector is
 \bea\label{fRe4}
U_{{\rm R-R}} = U_{C_{p+1}} + U_{C_{p-3}}= -8n_1n_2\left(1+f_1^2
f_2^2\right)  c_p^2 \frac{V_{p+1}}{k_\bot^2}.
 \eea

From Eqs. (\ref{fNSe4}) and (\ref{fRe4}), we know the total energy
density from both sectors is
 \bea\label{fe4}
U = U_{{\rm NS-NS}} + U_{{\rm R-R}} = 0.
 \eea
This shows that the interaction between two (D$_{p-4}$, D$_p$)
vanishes. This can be understood from the well-known fact that there
is no interaction between two constituent D$_p$-branes or between
two constituent D$_{p - 4}$-branes or between one D$_p$-brane and
one D$_{p - 4}$-brane. This net-zero interaction indicates that the
underlying system preserves also 1/4 of the spacetime
supersymmetries.

\begin{itemize}

\item The long-range interaction between (D$_{p-4}$, D$_p$) and ($\bar{\rm D}_{p-4}$, D$_p$)

The charge of a $\bar{\rm D}$-brane, i.e., anti D-brane, has the
opposite sign to that of a D-brane. The configuration of ($\bar{\rm
D}_{p-4}$, D$_p$) can be obtained with the following flux $\hat F$
 \bea\label{flux4anti}
\hat F =
\begin{pmatrix}  0_{2n} & \, & \, &\, &\, &\, &\,&\, \\
\, & \cdot & \, &\, &\, &\, &\,&\,\\
\, & \, & \cdot &\, &\, &\, &\,&\,\\
\, & \, & \, &\cdot &\, &\, &\,&\,\\
\, & \, & \, &\, &0_{2n} &f\cdot u &\,&\,\\
\, & \, & \, &\, &- f\cdot u &0_{2n}  &\,&\,\\
\, & \, & \, &\, &\, &\, &0_{2n} &-f\cdot u\\
\, & \, & \, &\, &\, &\, &f\cdot u
&0_{2n}\end{pmatrix}_{(p+1)\times(p+1)}.
 \eea
With this flux, $J_h$, $J_\phi$, $J_B$ and $J_{C_{p+1}}$ remain the
same as before while $J_{C_{p-3}}$ changes its sign. So the total
energy density is
 \bea\label{fe4antipm4}
U = U_h + U_\phi  + U_{C_{p+1}} - U_{C_{p-3}}=16n_1n_2f_1^2f_2^2
c_p^2 \frac{V_{p+1}}{k_\bot^2},
 \eea
which implies  the expected attractive interaction between
(D$_{p-4}$, D$_p$) and ($\bar{\rm D}_{p-4}$, D$_p$). This attractive
interaction is actually due to that between D$_{p-4}$-branes and
$\bar{\rm D}_{p-4}$-branes.

\item The long-range interaction between (D$_{p-4}$, D$_p$) and (D$_{p-4}$, $\bar{\rm D}_p$)

The configuration (D$_{p-4}$, $\bar{\rm D}_p$) can be realized via
the flux $\hat F$ given in Eq. (\ref{flux4anti}) and with the change
of sign of the corresponding Wess-Zumino term in Eq. (\ref{fwv}).
With these, $J_h$, $J_\phi$, $J_B$ and $J_{C_{p-3}}$ don't change
while $J_{C_{p+1}}$ changes sign. So the total energy density is
 \bea\label{fe4antip}
U = U_h + U_\phi - U_{C_{p+1}} + U_{C_{p-3}}=16n_1n_2 c_p^2
\frac{V_{p+1}}{k_\bot^2},
 \eea
which again implies an attractive interaction between (D$_{p-4}$,
D$_p$) and (D$_{p-4}$, $\bar{\rm D}_p$). It is now due to that
between D$_p$-branes and $\bar{\rm D}_p$-branes.

\item The long-range interaction between (D$_{p-4}$, D$_p$) and ($\bar{\rm D}_{p-4}$, $\bar{\rm D}_p$)

The state ($\bar{\rm D}_{p-4}$, $\bar{\rm D}_p$) can be obtained
simply by the change of sign of the corresponding Wess-Zumino term
in Eq. (\ref{fwv}). With this, $J_h$, $J_\phi$ and $J_B$ remain the
same as before while $J_{C_{p+1}}$ and $J_{C_{p-3}}$ both change
their signs. So the total energy density is
 \bea\label{fe4anti} U
&=& U_h + U_\phi - U_{C_{p+1}} - U_{C_{p-3}}\nn
&=&16n_1n_2(1+f_1^2f_2^2)  c_p^2 \frac{V_{p+1}}{k_\bot^2},
 \eea
which is the sum of Eqs. (\ref{fe4antipm4}) and (\ref{fe4antip}) as
expected. This indicates that this net interaction is due to that
between D$_p$-branes and $\bar{\rm D}_p$-branes and that between
D$_{p-4}$-branes and $\bar{\rm D}_{p-4}$-branes but there is no
interaction between D$_p$-branes and ${\bar D}_{p - 4}$-branes or
between D$_{p - 4}$-branes and ${\bar D}_p$-branes, again as
expected. This property is just a consequence of (D$_{p - 4}$,
D$_p$) as a marginal bound state.

\end{itemize}

\subsection{The $k = 3$ case}

The calculations of the long-range interaction for this case follow
basically the same steps as in the previous subsection. The
configuration of D$_{p-6}$-branes within D$_p$-branes can be
realized with the following constant non-abelian magnetic flux $\hat
F$
 \bea\label{flux6}
\hat F =
\begin{pmatrix} 0_{4n} & \, & \, &\, &\, &\, &\,&\,&\,&\, \\
\, & \cdot & \, &\, &\, &\, &\,&\,&\,&\,\\
\, & \, & \cdot &\, &\, &\, &\,&\,&\,&\,\\
\, & \, & \, &\cdot &\, &\, &\,&\,&\,&\,\\
\, & \, & \, &\, &0_{4n} &-f\cdot u_1 &\,&\,&\,&\,\\
\, & \, & \, &\, &f\cdot u_1 &0_{4n}  &\,&\,&\,&\,\\
\, & \, & \, &\, &\, &\, &0_{4n} &-f\cdot u_2 &\,&\,\\
\, & \, & \, &\, &\, &\, &f\cdot u_2 &0_{4n} &\,&\, \\
\, & \, & \, &\, &\, &\, &\,&\, &0_{4n} &-f\cdot u_3 \\
\, & \, & \, &\, &\, &\, &\,&\, &f\cdot u_3 &0_{4n}
\end{pmatrix}_{(p+1)\times(p+1)},
 \eea
where the $0_{4n}$ stands for $4n\times 4n$ zero matrix, and
 \bea\label{u6}
u_1 &=& {\rm Diag}\{I_n,I_n,-I_n,-I_n\}, \nn u_2 &=& {\rm
Diag}\{I_n,-I_n,-I_n,I_n\}, \nn u_3 &=& {\rm
Diag}\{I_n,-I_n,I_n,-I_n\}.
 \eea
Note that $u_1$, $u_2$ and $u_3$ are three of the Cartan subalgebra
generators of the $U(N)$ algebra with now  $N=4n$ ($n$ is a positive
integer). With this flux, ${\rm Tr} (\hat F\wedge \cdots \wedge \hat
F) \neq 0$ only when the number of $\hat F$ is 3 in the wedge
product. Therefore from Eq. (\ref{fcpmc}), we know that the only
non-vanishing coupling with the lower rank R-R potential is for $k =
3$ and the corresponding R-R potential is $C_{p-5}$. This implies
that  we have D$_{p-6}$-branes within D$_p$-branes with their charge
proportional to the integral of ${\rm Tr} (\hat F\wedge \hat F
\wedge \hat F)$. This brane configuration whose energy exceeds the
sum of the energy of D$_{p-6}$-branes and D$_p$-branes is not a
bound state in the usual sense but a relative stable state, and
breaks all spacetime supersymmetries \cite{Taylor:1997aa}. We will
denote this configuration as (D$_{p-6}$, D$_p$) in the following.

With the above flux, by the same token, we have the following
couplings as
 \bea\label{fNS6c}
J_h^{(i)} &=& -4n_i c_p V_{p+1}(1+f_i^2)^{3/2}\tilde
V_i^{\a\b}h_{\a\b}, \nn J_\phi^{(i)} &=& -4n_i \frac{c_p}{2\sqrt 2}
V_{p+1}
(1+f_i^2)^{3/2}\left[\left(p-3\right)-6\frac{f_i^2}{1+f_i^2}\right]\phi,
\nn J_B^{(i)} &=& -4n_i \frac{c_p}{\sqrt 2}
V_{p+1}(1+f_i^2)^{3/2}\tilde V_i^{\a\b}B_{\b\a}
 \eea
for the NS-NS fields and
 \bea\label{fR6c}
J_{C_{p+1}}^{(i)} &=& 4n_i \sqrt 2 c_p V_{p+1}C_{01\cdots p}, \nn
J_{C_{p-5}}^{(i)} &=& - 4n_i \sqrt 2 c_p V_{p+1} f_i^3 C_{01\cdots
p-6}
 \eea
for the R-R fields. In the above, we have the diagonal matrix
 \bea\label{tV6}
\tilde V_i ={\rm
Diag}\{-1,\,1,\,\cdots,\,1,\,\frac{1}{1+f^2_i},\,\frac{1}{1+f^2_i},
\,\frac{1}{1+f^2_i},\,\frac{1}{1+f^2_i},\,\frac{1}{1+f^2_i},\,\frac{1}{1+f^2_i}\},
 \eea
which immediately implies $J^{(i)}_B = 0$.
 Using Eqs. (\ref{fNS6c}) -
(\ref{tV6}), we then have the respective long-range interaction
energy density in momentum space due to the exchange of  the
corresponding massless field as
 \bea\label{fNSse6}
U_h &=& c_p^2 \frac{V_{p+1}}{k_\bot^2} 16n_1n_2 \sqrt
{\left(1+f_1^2\right)\left(1+f_2^2\right)} \left(
\frac{-p^2+6p+7}{8} +\frac{-p^2+12p-35}{8}f_1^2 \right.\nn &
&\left.\,\,\,\,\,\,\,\,\,\,\,\,\,\,\,\,+ \frac{-p^2+12p-35}{8}f_2^2+
\frac{-p^2+18p-65}{8}f_1^2 f_2^2\right), \nn U_\phi &=& c_p^2
\frac{V_{p+1}}{k_\bot^2} 16n_1n_2 \sqrt
{\left(1+f_1^2\right)\left(1+f_2^2\right)} \left( \frac{p^2-6p+9}{8}
+\frac{p^2-12p+27}{8}f_1^2 \right.\nn &
&\left.\,\,\,\,\,\,\,\,\,\,\,\,\,\,\,\,+ \frac{p^2-12p+27}{8}f_2^2+
\frac{p^2-18p+81}{8}f_1^2 f_2^2\right), \nn U_B &=& 0
 \eea
for the NS-NS fields and
 \bea\label{fRs6}
U_{C_{p+1}} &=& -32 n_1n_2 c_p^2 \frac{V_{p+1}}{k_\bot^2} , \nn
U_{C_{p-5}} &=& -32 n_1n_2 f_1^3f_2^3  c_p^2
\frac{V_{p+1}}{k_\bot^2}
 \eea
for the R-R fields. The total contribution to the energy density
from the NS-NS sector is
 \bea\label{fNSe6}
U_{{\rm NS-NS}} &=& U_h + U_\phi + U_B \nn &=& 16n_1n_2 \sqrt
{\left(1+f_1^2\right)\left(1+f_2^2\right)}
\left(2-f_1^2-f_2^2+2f_1^2f_2^2\right) c_p^2
\frac{V_{p+1}}{k_\bot^2} ,
 \eea
while the total one from the R-R sector is
 \bea\label{fRe6}
U_{{\rm R-R}} = U_{C_{p+1}} + U_{C_{p-5}}= -32n_1n_2\left(1+f_1^3
f_2^3\right)  c_p^2 \frac{V_{p+1}}{k_\bot^2}.
 \eea
So the total energy density from both sectors is
 \bea\label{fe6}
U &=& U_{{\rm NS-NS}} + U_{{\rm R-R}} \nn &=& c_p^2
\frac{V_{p+1}}{k_\bot^2} 16n_1n_2 \left[\sqrt
{\left(1+f_1^2\right)\left(1+f_2^2\right)}
\left(2-f_1^2-f_2^2+2f_1^2f_2^2\right)-2\left(1+f_1^3f_2^3\right)\right].~~
 \eea
We can show \footnote{If $2 (1 + f_1^2 f_2^2) \le f_1^2 + f_2^2$,
this is obviously true. So we need to check that this remains so for
$2 (1 + f_1^2 f_2^2) > f_1^2 + f_2^2$ with $f_1 f_2 \ge 0$. For
this, we need to show that $\sqrt{(1 + f_1^2)(1 + f_2^2)} \left[ 2(1
+ f_1^2 f_2^2) - f_1^2 - f_2^2\right] \leq 2 (1 + f_1^3 f_2^3)$
which is equivalent to  $(1 + f_1^2)(1 + f_2^2) \left[ 2(1 + f_1^2
f_2^2) - f_1^2 - f_2^2\right]^2 \leq 4 (1 + f_1^3 f_2^3)^2$ since
the left and right of the inequality are both positive. This latter
inequality can be simplified to $(f_1 - f_2)^2
\left[\left(3+3f_1^2f_2^2-f_1^2-f_2^2\right)\left(f_1+f_2\right)^2-4f_1^2f_2^2
\right]\geq 0$. Note that $\left(f_1+f_2\right)^2 \geq 4f_1f_2$,
then we have the term in the square bracket greater than or equal to
zero if $f_1 f_2 \ge 0$ . Therefore we have $U \le 0$ if $f_1 f_2
\ge 0$. } that $U\leq0$ if $f_1 f_2 \ge 0$ and the equality holds
only if $f_1 = f_2$.  This indicates that the interaction between
two (D$_{p-6}$, D$_p$) is in general repulsive and vanishes only if
the two fluxes are identical. Note that, as pointed out in
\cite{polbooktwo}, the (D$_{p-6}$, D$_p$) system itself doesn't
preserve any supersymmetry and is unstable, so the $U = 0$ case
doesn't imply any supersymmetry preservation of the interacting
system under consideration, unlike the other cases such as the
non-threshold BPS (D$_{p-2}$, D$_p$) bound states.

By the same token, the force nature for other cases as in $k = 2$
can also be analyzed and discussed.

\subsection{The $k = 4$ case}

This case corresponds to the configuration of D$_{p-8}$-branes
within D$_p$-branes. The only relevant case is for $p = 8$, i.e.,
D$_0$-branes within D$_8$-branes.  The brane system  can be realized
with D$_8$-branes carrying the following constant non-abelian
magnetic flux $\hat F$
 \bea\label{flux8}
\hat F =
\begin{pmatrix} 0_{8n} & \, & \, &\, &\, &\, &\,&\,&\, \\
\, & 0_{8n} &-f\cdot u_1 &\, &\, &\,&\,&\,&\,\\
\, & -f\cdot u_1 &0_{8n} &\, &\, &\,&\,&\,&\,\\
\, & \, &\, &0_{8n} &-f\cdot u_2 &\,&\,&\,&\,\\
\, & \, &\, &f\cdot u_2 &0_{8n}  &\,&\,&\,&\,\\
\, & \, &\, &\, &\, &0_{8n} &-f\cdot u_3 &\,&\,\\
\, & \, &\, &\, &\, &f\cdot u_3 &0_{8n} &\,&\, \\
\, & \, &\, &\, &\, &\,&\, &0_{8n} &-f\cdot u_4 \\
\, & \, &\, &\, &\, &\,&\, &f\cdot u_4 &0_{8n}
\end{pmatrix}_{9\times9},
 \eea
where the $0_{8n}$ stands for $8n\times 8n$ zero matrix, and
 \bea\label{u8}
u_1 &=& {\rm Diag}\{I_n,I_n,I_n,I_n,-I_n,-I_n,-I_n,-I_n\}, \nn u_2
&=& {\rm Diag}\{I_n,I_n,-I_n,-I_n,I_n,I_n,-I_n,-I_n\}, \nn u_3 &=&
{\rm Diag}\{I_n,-I_n,I_n,-I_n,I_n,-I_n,I_n,-I_n\}, \nn u_4 &=& {\rm
Diag}\{I_n,-I_n,-I_n,I_n,-I_n,I_n,I_n,-I_n\}.
 \eea
Note that $u_1$, $u_2$, $u_3$ and $u_4$ are four of the Cartan
subalgebra generators of the $U(N)$ algebra with now  $N=8n$ ($n$ is
a positive integer). With this flux, ${\rm Tr} (\hat F\wedge \cdots
\wedge \hat F) \neq 0$ only when the number of $\hat F$ is 4 in the
wedge product. Therefore from Eq. (\ref{fcpmc}), we know that the
only non-vanishing coupling with the lower rank R-R potential is for
$k = 4$ and the corresponding R-R potential is $C_1$. This implies
that we have D$_0$-branes within D$_8$-branes with the D$_0$-brane
charge proportional to the integral of ${\rm Tr} (\hat F\wedge \hat
F \wedge \hat F \wedge \hat F)$. We will denote this configuration
as (D$_0$, D$_8$) in the following.

With the above flux, by the same token, we have the following
couplings \footnote{ Note that in what follows for convenience we
express most of quantities in terms of $p$ unless explicitly
specified but it should be understood that $p = 8$ always.} as
 \bea\label{fNS8c}
J_h^{(i)} &=& -8 n_i c_p V_{p + 1} (1+f_i^2)^2\tilde
V_i^{\a\b}h_{\a\b}, \nn J_\phi^{(i)} &=& -8 n_i \frac{c_p}{2\sqrt 2}
V_{p + 1} (1+f_i^2)^2\left[ (p - 3) -
8\frac{f_i^2}{1+f_i^2}\right]\phi, \nn J_B^{(i)} &=& -8 n_i
\frac{c_p}{\sqrt 2} V_{p + 1} (1+f_i^2)^2\tilde V_i^{\a\b}B_{\b\a}
 \eea
for the NS-NS fields,  and
 \bea\label{fR8c}
J_{C_{p + 1}}^{(i)} &=& 8 n_i \sqrt 2 c_p V_{p + 1} C_{01\cdots 8},
\nn J_{C_{p - 7}}^{(i)} &=&  8n_i \sqrt 2 c_p V_{p + 1} f_i^4 C_0
 \eea
for the R-R fields. In the above, we have the diagonal matrix
 \bea\label{tV8}
\tilde V_i ={\rm
Diag}\{-1,\,\frac{1}{1+f^2_i},\,\frac{1}{1+f^2_i},\,\frac{1}{1+f^2_i},\,\frac{1}{1+f^2_i},
\,\frac{1}{1+f^2_i},\,\frac{1}{1+f^2_i},\,\frac{1}{1+f^2_i},\,\frac{1}{1+f^2_i}\},
 \eea
which immediately implies $J^{(i)}_B = 0$. Similarly, using Eqs.
(\ref{fNS8c}) - (\ref{tV8}), we then have the respective long-range
interaction energy density in momentum space due to the exchange of
the corresponding massless field as
 \bea\label{fNSse8}
U_h &=& c_p^2 \frac{V_{p+1}}{k_\bot^2} 64n_1n_2
\left(1+f_1^2\right)\left(1+f_2^2\right) \left( \frac{-p^2+6p+7}{8}
+\frac{-p^2+14p-49}{8}f_1^2 \right.\nn &
&\left.\,\,\,\,\,\,\,\,\,\,\,\,\,\,\,\,+ \frac{-p^2+14p-49}{8}f_2^2+
\frac{-p^2+22p-105}{8}f_1^2 f_2^2\right), \nn U_\phi &=& c_p^2
\frac{V_{p+1}}{k_\bot^2} 64n_1n_2
\left(1+f_1^2\right)\left(1+f_2^2\right) \left( \frac{p^2-6p+9}{8}
+\frac{p^2-14p+33}{8}f_1^2 \right.\nn &
&\left.\,\,\,\,\,\,\,\,\,\,\,\,\,\,\,\,+ \frac{p^2-14p+33}{8}f_2^2+
\frac{p^2-22p+121}{8}f_1^2 f_2^2\right), \nn U_B &=& 0
 \eea
for the NS-NS fields and
 \bea\label{fRs8}
U_{C_{p+1}} &=& -128 n_1n_2 c_p^2 \frac{V_{p+1}}{k_\bot^2} , \nn
U_{C_{p-7}} &=& -128 n_1n_2 f_1^4f_2^4  c_p^2
\frac{V_{p+1}}{k_\bot^2}
 \eea
for the R-R fields. For this particular system, there is an
additional coupling in the R-R sector between the one-form potential
$C_1$ and the nine-form potential $C_9$ because of the duality
relation for their components as $C_0 = - C_{01\cdots 8}$
\cite{Billo:1998np, Lu:2009w}. This coupling can also be interpreted
as arising from the half-string creation between a D$_0$-brane and a D$_8$-brane
\cite{Danielsson:1997wq, Bergman:1997gf, Kitao:1998vn} in the
present context as we will demonstrate in the following. The
corresponding contribution to the energy density can be calculated
as
 \bea\label{fRs8a}
U_{C_{p + 1}/C_{ p-7}} &=& \frac{1}{V_{p + 1}}
\left(\underbrace{J_{C_{p + 1}}^{(1)}J_{C_{p - 7}}^{(2)}} +
\underbrace{J_{C_{p - 7}}^{(1)}J_{C_{p + 1}}^{(2)}}  \right) \nn &=&
128 c_p^2 V_{p + 1}  n_1 n_2 \left(f_1^4+f_2^4
\right)\underbrace{C_{01\cdots 8} C_0}\nn &=& 128 n_1 n_2
\left(f_1^4+f_2^4\right) c_p^2 \frac{V_{p + 1}}{k_\bot^2},
 \eea
where we have used the above mentioned duality to give
 \bea\label{cpm7p}
\underbrace {C_{01\cdots 8}C_0} = -\underbrace {C_0
C_0}=-\underbrace {C_{01\cdots 8} C_{01\cdots 8}} =
\frac{1}{k_\bot^2}.
 \eea
In the above, we actually have two pieces with each positive and
coming from either the coupling between $J_{C_{p + 1}}^{(1)}$ and
$J_{C_{p - 7}}^{(2)}$ or  between $J_{C_{p -7}}^{(1)}$ and $J_{C_{p
+1}}^{(2)}$, therefore implying an attractive contribution.  For the
above mentioned purpose, we obtain the corresponding interaction in
coordinate space using Fourier transformation as \bea\label{Ampcoor}
U_{C_{p + 1}/C_{ p-7}}(Y)=\int \frac{d^\bot
k_\bot}{(2\pi)^\bot}e^{-\ii k_\bot \cdot Y}U_{C_{p + 1}/C_{
p-7}}=128 n_1 n_2 \left(f_1^4+f_2^4\right) c_p^2 V_{p + 1}
\left(I_{\infty}-\frac{Y}{2}\right).
 \eea
In the above, we have used the following relation
 \bea\label{transf}
\int \frac{d^\bot k_\bot}{(2\pi)^\bot}\frac{e^{-\ii k_\bot \cdot
Y}}{k_\bot^2}=I_{\infty}-\frac{Y}{2}
 \eea
for one transverse direction. Note that the $I_{\infty}$ in Eq.
(\ref{Ampcoor}) is positively infinity and independent of the
separation $Y$, representing the energy when $Y = 0$, and its
divergence actually reflects the D$_8$-brane nature of non-existence
as an independent object as discussed in \cite{Lu:2009w}. The
corresponding attractive force acting on the D$_0$-branes per unit
D$_0$-brane world-volume can be obtained as
 \bea\label{force}
F_{C_{p + 1}/C_{ p-7}}=-\frac{1}{V_{p-7}}\frac{d U_{C_{p + 1}/C_{
p-7}}(Y)}{d Y} = 64 n_1 n_2 \left(f_1^4+f_2^4\right) c_p^2
\frac{V_{p + 1}}{V_{p -7}}.
 \eea
For clearly demonstrating the half-string creation, we express the
above force as \be \label{forcepiece} F_{C_{p + 1}/C_{ p-7}} =
F^{(1)}_{C_{p + 1}/C_{ p-7}} + F^{(2)}_{C_{p + 1}/C_{ p-7}}\ee with
\be \label{pieceforce} F^{(i)}_{C_{p + 1}/C_{ p-7}} = 64 n_1 n_2
f_i^4 c_p^2 \frac{V_{p + 1}}{V_{p -7}},\ee where $i = 1$ or $2$.
From Eq. (\ref{fR8c}) or the integral of ${\rm Tr} (\hat F\wedge
\hat F \wedge \hat F \wedge \hat F)$, we have the quantization
condition
 \bea\label{qc}
8n_i f_i^4 c_p V_{p + 1}  = m_i c_{p-8}V_{p -7},
 \eea
where $m_i$ is an integer and represents the total number of
D$_0$-branes within D$_8$-branes. We have then, taking $i = 1$ for
example,
 \bea\label{force1}
F^{(1)}_{C_{p + 1}/C_{p-7}} = 8 m_1 n_2  c_0 c_8 = 8  m_1 n_2
\frac{1}{4\pi\a'},
 \eea
where we have used the exact value of $c_p$ for D$_p$-brane as
$c_p=\sqrt{\pi}\left(2\pi\sqrt{\a'}\right)^{3-p}$
\cite{Billo:1998np, Lu:2009nwx}. Note that $N_2 = 8 n_2$  is the
total number of D$_8$-branes in the second (D$_0$, D$_8$) bound
state and the $m_1$ is the total number of D$_0$-branes in the first
bound state, therefore the force between a D$_0$-brane and a
D$_8$-brane is given from the above as \be \frac{F^{(1)}_{C_{p +
1}/C_{p-7}}}{8 m_1 n_2} = \frac{1}{4\pi\a'} = \frac{T}{2}\ee where
$T$ is the tension of a fundamental string. This demonstrates that a
string with its tension one half of a fundamental string is created
between a D$_0$-brane and a D$_8$-brane, following the same spirit
of \cite{Danielsson:1997wq, Bergman:1997gf, Kitao:1998vn, Lu:2009w}.
This interpretation is  in line with \cite{Polchinski:1995sm} for a
D$_0$-brane in the presence of a D$_8$-brane and is also consistent
with the Hanany-Witten effect \cite{Hanany:1996ie} for a D$_0$-brane
crossing a D$_8$-brane. The same is true if $i = 2$ is taken in the
above. So we demonstrate that a string with its tension one half of
a fundamental string is created between a D$_0$-brane and a
D$_8$-brane even in the non-abelian context, lending further support
to this assertion.

The total contribution to the energy density from the NS-NS sector
is
 \bea\label{fNSe8}
U_{{\rm NS-NS}} = U_h + U_\phi + U_B = 128n_1n_2
 \left(1 - f_1^4 - f_2^4 +
f_1^4 f_2^4\right) c_p^2 \frac{V_{p+1}}{k_\bot^2} ,
 \eea
while the total one from the R-R sector is
 \bea\label{fRe8}
U_{{\rm R-R}} = U_{C_{p+1}} + U_{C_{p-7}} + U_{C_{p+1}/C_{p-7}}=
-128n_1n_2\left(1 - f_1^4 - f_2^4 + f_1^4 f_2^4\right)  c_p^2
\frac{V_{p+1}}{k_\bot^2}.
 \eea
So the total energy density from both sectors is
 \bea\label{fe8}
U = U_{{\rm NS-NS}} + U_{{\rm R-R}} = 0.
 \eea
If the (D$_0$, D$_8$) is as expected a marginal bound state,  the
above indicates that the underlying system preserves 1/4 of
spacetime supersymmetries. Note that the contribution
$U_{C_{p+1}/C_{p-7}}$ is independent of the nature of the
constituent branes, i.e., being branes or anti-branes, in the
respective bound state. But this is not the case for the $U_{C_{p +
1}}$ or $U_{C_{p - 7}}$. In analogue to the case of (D$_{p-4}$,
D$_p$), we can analyze the following three cases.

\begin{itemize}

\item The long-range interaction between (D$_0$, D$_8$) and ($\bar{\rm D}_0$, D$_8$)

The presence of $\bar{\rm D}_0$-branes changes the sign of
$U_{C_{p-7}}$, so the total energy density is
 \bea\label{fe8antipm8}
U = U_h + U_\phi  + U_{C_{p+1}} - U_{C_{p-7}} + U_{C_{p+1}/C_{p-7}}
=256n_1n_2f_1^4f_2^4 c_p^2 \frac{V_{p+1}}{k_\bot^2},
 \eea
which implies  the expected attractive interaction between (D$_0$,
D$_8$) and ($\bar{\rm D}_0$, D$_8$). This attractive interaction is
actually due to that between D$_0$-branes and $\bar{\rm
D}_0$-branes.

\item The long-range interaction between (D$_0$, D$_8$) and (D$_0$, $\bar {\rm D}_8$)

The presence of $\bar{\rm D}_8$-branes changes the sign of
$U_{C_{p+1}}$, so the total energy density is
 \bea\label{fe8antip}
U = U_h + U_\phi  - U_{C_{p+1}} + U_{C_{p-7}} + U_{C_{p+1}/C_{p-7}}
=256n_1n_2 c_p^2 \frac{V_{p+1}}{k_\bot^2},
 \eea
which also implies an attractive interaction due to that between
D$_8$-branes and $\bar{\rm D}_8$-branes.

\item The long-range interaction between (D$_0$, D$_8$)
and ($\bar{\rm D}_0$, $\bar{\rm D}_8$)

The presence of $\bar{\rm D}_0$-branes and $\bar{\rm D}_8$-branes
changes the signs of $U_{C_{p-7}}$ and $U_{C_{p+1}}$, so the total
energy density is
 \bea\label{fe8antipantipm8}
U = U_h + U_\phi  - U_{C_{p+1}} - U_{C_{p-7}} + U_{C_{p+1}/C_{p-7}}
=256n_1n_2\left(1+f_1^4f_2^4\right) c_p^2 \frac{V_{p+1}}{k_\bot^2},
 \eea
which is the sum of Eqs. (\ref{fe8antipm8}) and (\ref{fe8antip}) as
expected. This indicates that this net interaction is due to that
between D$_0$-branes and $\bar{\rm D}_0$-branes and that between
D$_8$-branes and $\bar{\rm D}_8$-branes.

\end{itemize}

The above computations further imply that, in analogue to the case
of (D$_{p-4}$, D$_p$), the interaction between (D$_0$, D$_8$) and
($\bar{\rm D}_0$, D$_8$), or between (D$_0$, D$_8$) and (D$_0$,
$\bar {\rm D}_8$), or between (D$_0$, D$_8$) and ($\bar{\rm D}_0$,
$\bar{\rm D}_8$), is always attractive. It is entirely due to the
interaction between D$_0$-branes and $\bar {\rm D}_0$-branes or
between D$_8$-branes and $\bar {\rm D}_8$-branes, or the sum of the
interaction between D$_0$-branes and $\bar {\rm D}_0$-branes and
that between D$_8$-branes and $\bar {\rm D}_8$-branes, in the
respective bound states.

The above computations strongly suggest that the bound state (D$_0$,
D$_8$) so constructed is a marginal one but it was argued in
\cite{Taylor:1997aa} based on the relative scaling of $\hat F^2$ and
$\hat F \wedge \hat F \wedge \hat F \wedge \hat F$ while keeping the
D$_0$-brane charge invariant and on the energy argument that it is
not. While this confusion cannot be settled down for the time being,
we would like to point out that the peculiar nature of D$_8$-brane
being unable to be an independent object by itself \cite{polbooktwo}
and the associated divergent self-energy of the D$_8$-brane may indicate that
the scaling and the energy arguments mentioned above may be a bit
too simple and further re-examination of this is needed.
Nevertheless, we have demonstrated that the half-string creation
picture still holds even in the present context, lending further
support to this.

\section{Summary}

We have calculated the long-range interactions between two simple
$p$-branes in diverse dimensions without any world-volume flux
turned on. We also compute the interaction either between two
(D$_{p-4}$, D$_p$) or between two (D$_{p-6}$, D$_p$) or between two
(D$_0$, D$_8$), respectively, where the lower dimensional branes in
the respective state can be represented by the corresponding special
world-volume non-abelian magnetic flux.

For the simple $p$-brane case, the  static net interaction vanishes.
So the ``no-force'' condition holds and we have preserved 1/2 of the
spacetime supersymmetries. If we replace one set of coincidental
branes by its corresponding anti-branes, all supersymmetries are
broken and the long-range interaction between one set of
coincidental branes and one set of coincidental  anti-branes with a
separation is attractive and is explicitly given. For the case of
IIA NS5-brane or (transverse) M5-brane, this is the interaction
which may only be computed at the present.

 For (D$_{p-4}$, D$_p$),
i.e., a marginally bound state, our calculations confirm the
well-known fact that the interaction between two such brane
configurations is the sum of two contributions: one is due to the
two sets of D$_{p-4}$-branes in the two configurations,
respectively, and the other is due to the two sets of D$_p$-branes.
There is no contribution from the D$_{p-4}$-branes in one
configuration and the  D$_p$-branes in the other one. When one or
both the constituent branes in one bound state are taken as the
anti-ones, the corresponding explicit interaction potential computed
in this paper is believed to be given the first time.

The case of (D$_{p-6}$, D$_p$) is a bit more complicated. This
system itself doesn't preserve any supersymmetry and the interaction
between two such systems is in general repulsive just like that
between two constituents in the state (D$_{p-6}$, D$_p$). The
long-range interaction can still vanish but this doesn't imply any
preservation of supersymmetry for the interacting system under
consideration. Once again, the general explicit interaction
potential in this case is computed the first time.

For the case of (D$_0$, D$_8$). First the long-range interaction
between a D$_0$-brane and a D$_8$-brane can also be calculated by
the same token if a key duality relation between the R-R potential
$C_0$ associated with the D$_0$-brane and the R-R potential
$C_{01\cdots 8}$ associated with the D$_8$-brane, namely $C_0 = -
C_{01\cdots 8}$ found in \cite{Billo:1998np}, is employed. The
counter-intuitive R-R contribution was calculated via an effective
field approach by one of the present authors in \cite{Lu:2009w} and
the NS-NS contribution can be trivially calculated via the method
described in this paper. One peculiar feature of this system is that
only the massless modes rather than the full string spectrum
contribute to the lowest-order stringy interaction and therefore it
can be calculated via an effective field theory approach. We also
demonstrate that the half-string creation in the present context
continues to hold, therefore lending further support to the previous
assertion of this. Non-vanishing potentials for variants of such
system are computed explicitly.

Our computations indicate that the interaction between two (D$_0$,
D$_8$) follows the same line as the case of (D$_{p - 4}$, D$_p$),
therefore strongly suggesting that the bound state so constructed
with the special constant non-abelian magnetic flux is an expected
marginal bound state. This may indicate that a further
re-examination of the analysis given in \cite{Taylor:1997aa} for
this bound state is needed along with the consideration of the
D$_8$-brane nature that this co-dimensional one brane cannot exist
by itself as an independent object.

\vspace{.5cm}

\noindent {\bf Acknowledgements}

We would like to thank J. X. Lu for his help in improving  the
manuscript and for many very fruitful discussions. We acknowledge
support by grants from the Chinese Academy of Sciences, a grant from
973 Program with grant No: 2007CB815401 and grants from the NSF of
China with Grant No:10588503, 10535060 and 10975129.

\end{document}